\documentclass[aps,prl,onecolumn,superscriptaddress,groupedaddress]{revtex4}  
\usepackage{graphicx}  
\usepackage{color, soul}
\usepackage{xcolor}
\usepackage{dcolumn}   
\usepackage{bm}        
\usepackage{amssymb}   
\usepackage{changepage}
\usepackage{amsmath}
\usepackage{hyperref}
\usepackage{float}
\usepackage{blindtext}
\hypersetup{
  colorlinks=false,
  linkbordercolor=blue
}

\usepackage[a4paper,bindingoffset=0.1in,%
            left=1.5cm,right=1.5cm,top=2cm,bottom=2.5cm,%
            footskip=0.5in]{geometry}
\usepackage{blindtext}

\begin{document}


\newlength{\halfpagewidth}
\setlength{\halfpagewidth}{\linewidth}
\divide\halfpagewidth by 2
\newcommand{\leftsep}{%
\noindent\raisebox{4mm}[0ex][0ex]{%
\makebox[\halfpagewidth]{\hrulefill}\hbox{\vrule height 3pt}}%
}
\newcommand{\rightsep}{%
\noindent\hspace*{\halfpagewidth}%
\rlap{\raisebox{-3pt}[0ex][0ex]{\hbox{\vrule height 3pt}}}%
\makebox[\halfpagewidth]{\hrulefill} } 
    
\title{Thermal Transfer in Amorphous Superionic Systems }
\author{Yanguang~Zhou}
\email{maeygzhou@ust.hk}
\affiliation{Department of Mechanical and Aerospace Engineering, The Hong Kong University of Science and Technology, Clear Water Bay, Hong Kong, China}

\author{Sebastian~Volz}
\affiliation{CNRS, UPR 288 Laboratoire d’Energétique Moléculaire et Macroscopique, Combustion (EM2C), Ecole Centrale Paris, Grande Voie des Vignes, 92295, Châtenay-Malabry, France}    
\affiliation{LIMMS/CNRS-IIS(UMI2820) Institute of Industrial Science, University of Tokyo 4-6-1 Komaba, Meguro-ku Tokyo, 153-8505, Japan}     

\date{\today}

\begin{abstract}

Using direct atomic simulations, the vibration scattering time scales are characterized, and then the nature and the quantitative weight of thermal excitations are investigated in an example system Li$_{2}$S from its amorphous solid state to its partial-solid partial-liquid and, liquid states. For the amorphous solid state at 300 K, the vibration scattering time ranges a few femtoseconds to several picoseconds. As a result, both the progagons and diffusons are the main heat carriers and contribute largely to the total thermal conductivity. The enhancement of scattering among vibrations and between vibrations and free ions flow due to the increase of temperature, will lead to a large reduction of the scattering time scale and the acoustic vibrational thermal conductivity, i.e., 0.8 W/mK at 300 K to 0.56 W/mK in the partial solid partial liquid Li$_{2}$S at 700 K. In this latter state, the thermal conductivity contributed by convection increases to the half of the total, as a result of the usually neglected cross-correlation between the virial term and the free ions\rq\  flow. The vibrational scattering time can be as large as $\sim$ 1.5 picoseconds yet, and the vibrational conductivity is reduced to a still significant 0.42 W/mK highlighting the unexpected role of acoustic transverse and longitudinal vibrations in liquid Li$_{2}$S at 1100 K. At this same temperature, the convection heat transfer takes overreaching 0.63 W/mK. Our study provides a fundamental understanding of the thermal excitations at play in amorphous materials from solid to liquid. 

\end{abstract}
\setulcolor{blue}
\maketitle
\twocolumngrid 


Conduction and convection heat transfer is the main heat exchange modes in solids and liquids, respectively, which is of critical importance in energy conversion \cite{Lan2010, Bux2009}, energy storage \cite{Huang2019} and energy transport \cite{Cahill2003}. While it is well-known that phonons, i.e. an energy quantum of propagating lattice waves, are dominant thermal excitations in perfect crystals \cite{Kittel1976}, thermal excitations in disordered solids as well as in liquids are much more diverse and difficult to be quantitatively characterized. Thermal excitations in amorphous solids and nanostructures, e.g., amorphous Si \cite{Allen1993, Moon2018, Larkin2014, Zhou2015, Saaskilahti2016} and polycrystalline Si nanowires \cite{Zhou2016}, are classified as propagons, diffusons and locons depending on the degree of delocalization of atomistic vibrations and on their mean free path. This description was firstly proposed by Allen and Feldman \cite{Allen1993} and Cahill \cite{Cahill1992}, then extended by Larkin and MacGauhey \cite{Larkin2014} and Lv and Herny \cite{Lv2016}. The similar classification is also applied to quantify the thermal excitations in superionic systems which are in the partial-liquid partial-solid (PLPS) state at high temperatures such as crystal Li$_{2}$S \cite{Zhou2018}, Ag$_{2}$Te \cite{Wu2018}, AgCrSe$_{2}$ \cite{Li2018} and Cu$_{3}$SbSe$_{3}$ \cite{Qiu2014}, but with an extra contribution from convection heat transfer as proved by Zhou $et.$ $al.$ \cite{Zhou2018}. In liquids where convection heat transfer predominates, acoustic and optical phonons are experimentally observed such as in liquid Ga \cite{Hosokawa2009}, water \cite{Elton2016} and liquid Helium \cite{Woods1970}. Despite these progress in discovering the persistence of those modes in liquids, quantitatively characterizing the contribution of each thermal excitations in partial-solid partial-liquid and liquid systems remains challenging and poorly documented. 
   
In this letter, quantitative characterization of the vibrational scattering time scales and the contribution of each thermal mode from the amorphous solid to the liquid state transition in the example system of Li$_{2}$S was achieved by using atomistic simulations. The vibrational scattering time scale is found to be large and ranging from ranges a few femtoseconds to several picoseconds in Li$_{2}$S solids and liquids. Our analysis supports that conduction heat transfer contributed from propagons and diffusons is the dominant mode in amorphous Li$_{2}$S and contributes to around one half and one third to the total thermal energy exchange in PSPL and liquid Li$_{2}$S, respectively. Convection heat transfer which results from two different mechanisms, i.e., virial atomic interactions and ions\rq \ free movement, is found to increasingly contribute to the thermal energy exchange with the fluidification of Li and S ions. Furthermore, the conduction-convection interaction represents a non-negligible part of the total thermal conductivity in liquid Li$_{2}$S.

A Li$_{2}$S crystal including 6144 atoms \cite{SM2020} with interactions depicted by ReaxFF \cite{Islam2015} was firstly equilibrated at a high-temperature liquid state, i.e., 3000 K, for 62.5 ps and then cooled down to 300 K in the NPT (constant particle number, pressure and temperature) ensemble with a cooling rate of 10 K/ps. The structure is finally equilibrated at 300 K for another 62.5 ps in the NVT (constant particle number, volume and temperature) ensemble to reduce the metastabilities. Next, the generated amorphous Li$_{2}$S was annealed from 300 K to 1500 K for 375 ps in the NVT ensemble to simulate the amorphous, partial-solid partial-liquid and liquid Li$_{2}$S. 

The heat current which is used as an input in thermal conductivity calculations based on Green-Kubo theory \cite{Kubo1957} was computed for 12.5 ps in the NVE (constant particle number, volume and energy) ensemble. The time step in all the simulations is 0.25 fs. Periodic boundary conditions were applied in all three directions. 

To analyze the thermal transport behavior in the systems, we start from the expression of the heat current $\vec{{Q}''}$ \cite{Hardy1963}:
\begin{equation}
\begin{aligned}
&\vec{{Q}''}={{\vec{{Q}''}}_{convection}}+{{\vec{{Q}''}}_{virial}} =\underbrace{\sum\limits_{i}^{N}{{{e}_{i}}\cdot {{{\vec{v}}}_{i}}}}_{{{{{Q}''}}_{convection}}}+\\ 
&\underbrace{\frac{1}{2}\sum\limits_{\ i\ne j}^{N}{({{{\vec{F}}}_{ij}}\cdot {{{\vec{v}}}_{i}})\cdot {{{\vec{r}}}_{ij}}}+\sum\limits_{\ i\ne j\ne k}^{N}{({{{\vec{F}}}_{ijk}}\cdot \vec{v})({{{\vec{r}}}_{ij}}+{{{\vec{r}}}_{ik}})}}_{{{{{Q}''}}_{virial}}}
\end{aligned}
\label{eqn:A1}  
\end{equation}
in which ${{e}_{i}}$ refers to the ionic energy, ${{\vec{v}}_{i}}$ is the ionic velocity, ${{\vec{r}}_{ij}}$ the distance between two ions $i$ and $j$ and ${{\vec{F}}_{ij}}$ and ${{\vec{F}}_{ijk}}$ represent the two-body and three-body forces, respectively. The first term in Eq.\ (\ref{eqn:A1}) results from the free motion of ions, i.e. convection heat current ${{\vec{{Q}''}}_{convection}}$. The last two terms are corresponding to the interactions between two ions and are also called virial heat current ${{\vec{{Q}''}}_{virial}}$. Consequently, thermal conductivity $\kappa$ can be divided into contributions involving ion convection ${{\kappa }_{convection}}$, interaction among ions ${{\kappa }_{virial}}$, and a term resulting from convection-conduction interactions, ${{\kappa }_{cross}}$. Those three terms are derived from the Green-Kubo formula \cite{Kubo1957}:
\begin{equation}
\begin{aligned} 
 \kappa=& \ {{\kappa }_{virial}}+{{\kappa }_{convection}}+{{\kappa }_{cross}} \\ 
=&\ \frac{1}{3{{k}_{b}}V{{T}^{2}}}\int_{0}^{\infty }{\left[ \underbrace{\left\langle {{{\vec{{Q}''}}}_{virial}}(t)\cdot {{{\vec{{Q}''}}}_{virial}}(0) \right\rangle }_{virial} \right.}+ \\ 
 &\underbrace{\left\langle {{{\vec{{Q}''}}}_{convection}}(t)\cdot {{{\vec{{Q}''}}}_{convection}}(0) \right\rangle }_{convection}+ \\ 
 &\left. \underbrace{2\left\langle {{{\vec{{Q}''}}}_{virial}}(t)\cdot {{{\vec{{Q}''}}}_{convection}}(0) \right\rangle }_{cross} \right]dt 
\end{aligned}
\label{eqn:A2}  
\end{equation}
where ${{k}_{b}}$ is the Boltzmann constant, $V$ the system volume, $T$ represents the system temperature and $t$ denotes the autocorrelation time. The angular bracket indicates ensemble averaging. For each case, 30 independent runs are performed to obtain a stable averaged value of ${{\kappa }}$. The correlation time considered in our simulation is 12.5 ps, which is long enough to obtain the converged thermal conductivity \cite{SM2020}. Note that ${{\kappa }_{cross}}$ is not the result of scattering between lattice vibrations and liquid, but of the cross-correlation between the convection heat current and the virial one. Physically, the virial and convection heat currents depend on interatomic forces and are derived from standard statistical mechanics. Note that the cross term provides an additional virial heat current component generated by the convective heat current and vice-versa. 

\begin{figure}
\hspace*{-4mm} 
\setlength{\abovecaptionskip}{0.1in}
\setlength{\belowcaptionskip}{-0.1in}
\includegraphics [width=3.5in]{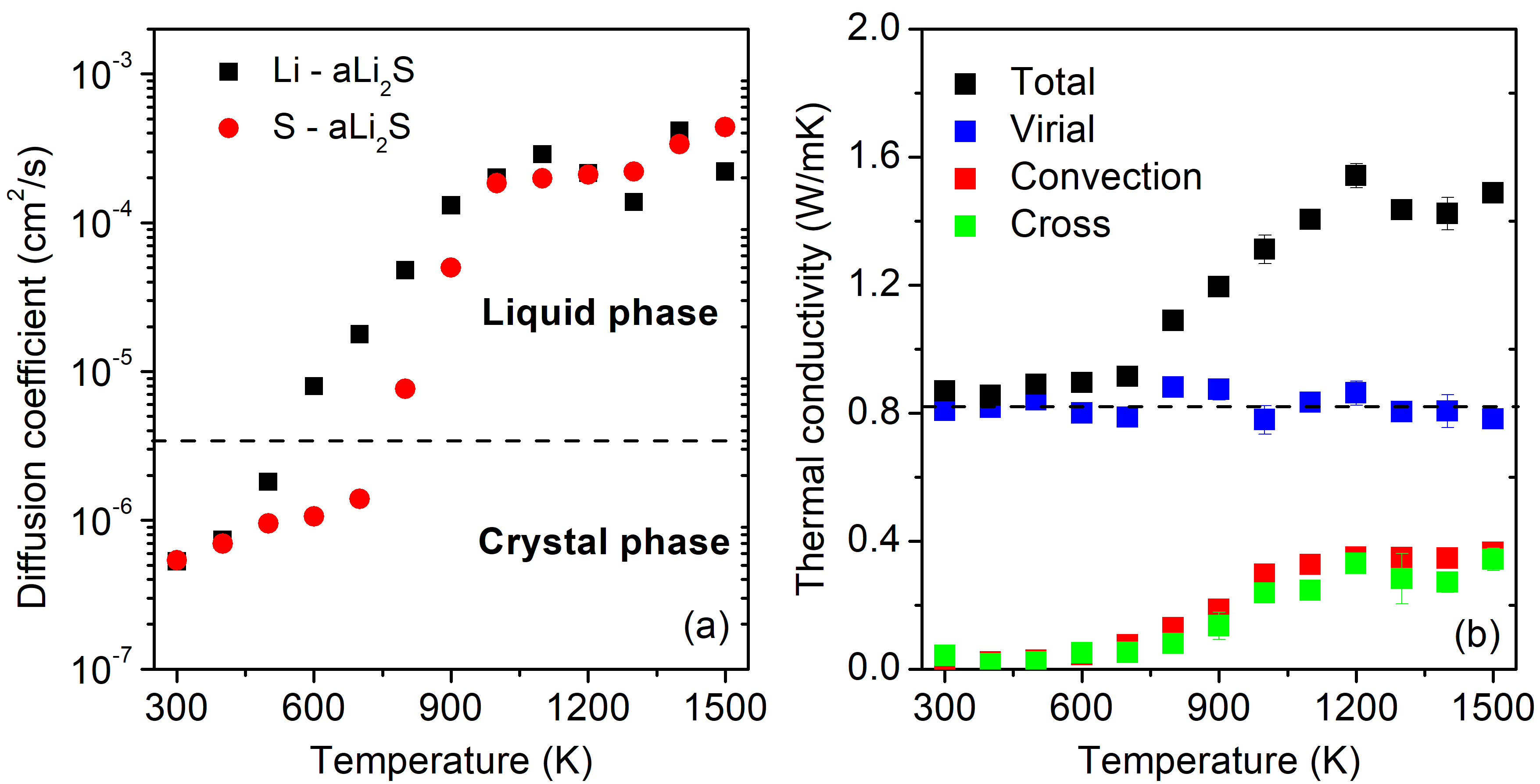}
\caption{(a) Diffusion coefficient of Li (black dots) and S (red dots) ions at different temperatures. (b) Thermal conductivity contributed from the virial term (blue dots), the convection term (red dots) and the cross term (green dots) computed using Eq.\ (\ref{eqn:A2}).}
\label{fig:F1}
\end{figure}

According to the diffusion coefficients of Li and S ions reported in Figure 1a, amorphous Li$_{2}$S is found to change from the solid-state to the partial solid partial liquid state at the transition temperature of 600 K and further to the liquid state at the critical temperature of 800 K (\textbf{Figure ~\ref{fig:F1}a}). Below 600 K, both Li and S ions manifest solid-like mobilities. When the temperature of the system is in the range of 600 K $\sim$ 800 K, S ions still remain in the solid-state while the mobility of Li ions is largely increased, which implies their fluidization. Above 800 K, both Li and S ions manifest liquid-like mobilities with high diffusion coefficients. Figure 1b shows the temperature-dependent thermal conductivity of the systems and the corresponding contributions from the virial atomic interactions, convection and cross terms. The thermal conductivity contributed by convection and cross terms is found to be negligible when amorphous Li$_{2}$S is in solid-state at temperatures below 600 K but to increase gradually in PSPL and liquid phases when the temperature is ranging from 600 K to 1100K. Thermal conductivity then further reaches a plateau due to the saturation of the diffusion coefficient (\textbf{Figure ~\ref{fig:F1}a}) when the system temperature exceeds 1100 K. The thermal conductivity resulting from the virial interactions among ions, which is usually regarded as conduction heat transfer, is almost independent of the temperature of the system. As a result, the total thermal conductivity has a similar temperature dependence as the thermal conductivity contributed by convection and cross terms. 

\begin{figure*}
\hspace*{-4mm} 
\setlength{\abovecaptionskip}{0.1in}
\setlength{\belowcaptionskip}{-0.1in}
\includegraphics [width=6in]{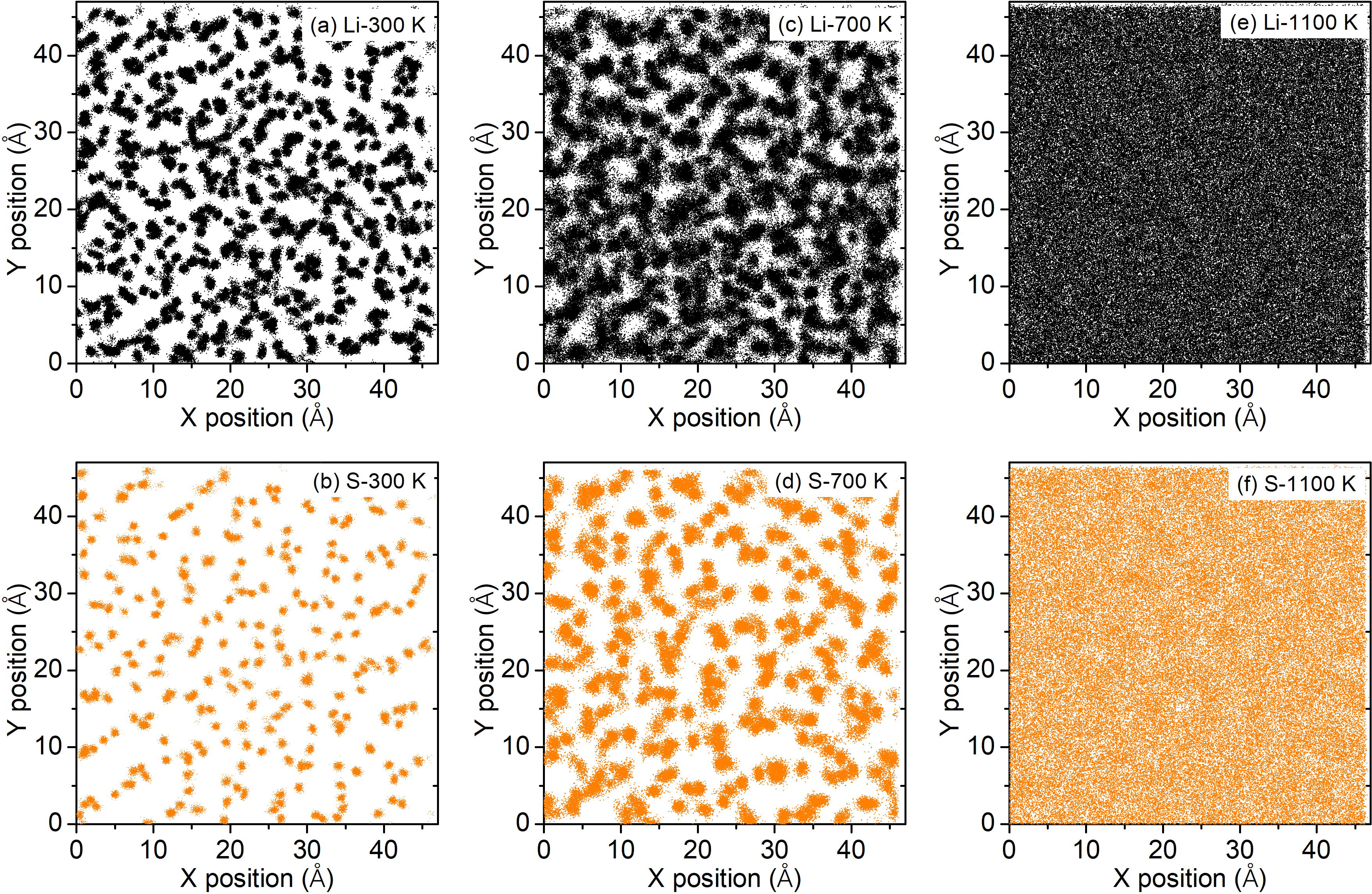}
\caption{Trajectories of Li ions (gray) and S ions (orange) at (a and b) 300 K, (c and d) 700 K and (e and f) 1100 K from the reactive force-field molecular dynamic simulations.}
\label{fig:F2}
\end{figure*}

To obtain an intuitive picture of the systems, we also trace the time trajectories of Li and S ions at three representative temperatures, i.e. 300 K, 700 K and 1100 K. Trajectories are projected onto the [100] lattice plane (\textbf{Figure ~\ref{fig:F2}}) as the superposition of many frames at different times uncovering the range in which ions can travel in real space. At low temperatures, i.e. 300 K, both Li and S ions vibrate around their equilibrium positions (\textbf{Figure ~\ref{fig:F2}a} and \textbf{\ref{fig:F2}b}). At a medium temperature of 700 K, although S ions still vibrate around their original sites, Li ions can travel far away from their equilibrium positions and flow inside the system (\textbf{Figure ~\ref{fig:F2}c} and \textbf{\ref{fig:F2}d}) indicating that they can be treated as in a liquid phase. At the high temperature of 1100 K, both Li and S ions can flow freely (\textbf{Figure ~\ref{fig:F2}e} and \textbf{\ref{fig:F2}f}) yielding liquid-like diffusion coefficients shown in \textbf{Figure ~\ref{fig:F1}}.                          

To gain more insight into the contribution to heat exchange from each thermal excitation, we calculate the vibrational dispersions in Li$_{2}$S using the dynamical structure factor $S$, given by \cite{Shintani2008}                  
\begin{equation}
\begin{aligned}
&{{S}_{T\ or\ L}}(|\vec{q}|,\ \omega )=\\  
&\int_{0}^{\infty }{\left\langle {{{\vec{\xi }}}_{T\ or\ L}}(|\vec{q}|,\ t)\cdot {{{\vec{\xi }}}_{T\ or\ L}}(-|\vec{q}|,\ 0) \right\rangle \exp (-i\omega t)dt}
\end{aligned}
\label{eqn:A3}  
\end{equation}
where $\vec{q}$ corresponds to the wave vector and $\omega$ to the angular frequency of the lattice vibrations. ${{\vec{\xi }}_{T}}$ and ${{\vec{\xi }}_{L}}$ stand for the ion velocities related to transverse and longitudinal polarizations, respectively, which can be computed via
\begin{equation}
{{\vec{\xi }}_{T}}(|\vec{q}|,\ t)=\sum{({{{\vec{v}}}_{i}}(t)-{{{\vec{v}}}_{i}}(t)\cdot \hat{\vec{q}})}\cdot \hat{\vec{q}}\exp (i\vec{q}\cdot {{\vec{r}}_{i}}(t))
\label{eqn:A4}  
\end{equation}
\begin{equation}
{{\vec{\xi }}_{L}}(|\vec{q}|,\ t)=\sum{({{{\vec{v}}}_{i}}(t)\cdot \hat{\vec{q}})}\cdot \hat{\vec{q}}\exp (i\vec{q}\cdot {{\vec{r}}_{i}}(t))
\label{eqn:A5}  
\end{equation}
in which $\hat{\vec{q}}=\vec{q}/|\vec{q}|$, and ${{\vec{r}}_{i}}$ refers to the position of the ion with index $i$. The structure factor has been successfully used to depict the dispersion relation in disorder solids and liquids \cite{Moon2018, Hosokawa2009, Shintani2008}. Following the discussion above, we calculate the dynamical structure factors of the three typical systems, i.e. amorphous solid Li$_{2}$S at 300 K, PSPL Li$_{2}$S at 700 K and liquid Li$_{2}$S at 1100 K. 

\begin{figure}[H]
\hspace*{-5mm} 
\setlength{\abovecaptionskip}{0.1in}
\setlength{\belowcaptionskip}{-0.1in}
\includegraphics [width=3.5in]{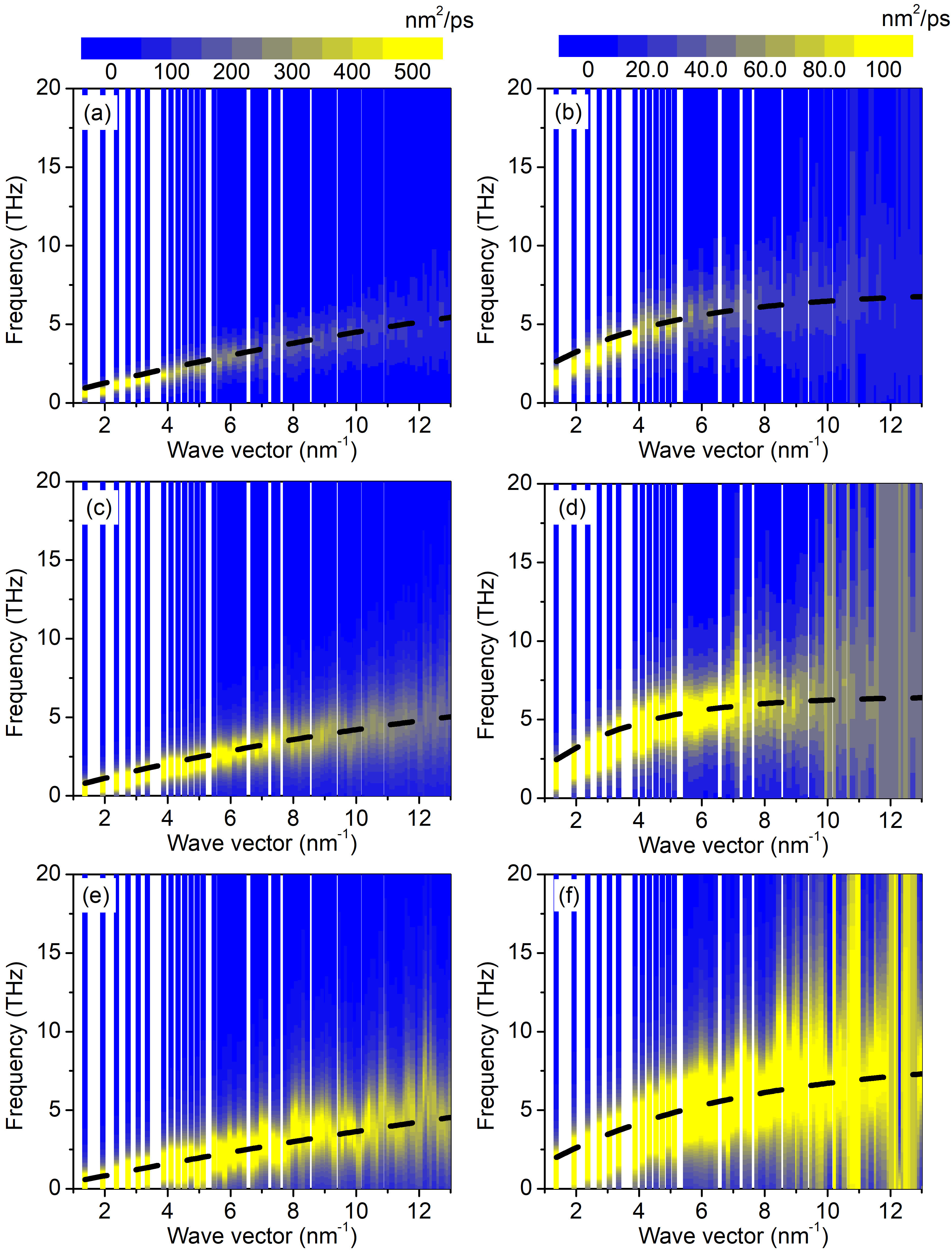}
\caption{Structure factors for the transverse branch at (a) 300 K, (c) 700 K and (e) 1100 K, and the longitudinal branch at (b) 300 K, (d) 700 K and (f) 1100 K. The black dashed lines are fitted dispersions.}
\label{fig:F3}
\end{figure}

As revealed by \textbf{Figure ~\ref{fig:F3}}, clearly defined dispersions exist up to 5 THz for the transverse branch and to 7.5 THz for the longitudinal one. At 300 K, since both Li and S ions have solid-like mobilities, transverse (\textbf{Figure ~\ref{fig:F3}a}) and longitudinal (\textbf{Figure ~\ref{fig:F3}b}) acoustic dispersions are observed in the systems with small frequency broadenings. By increasing the system temperature to 700 K (\textbf{Figure ~\ref{fig:F3}c} and \textbf{\ref{fig:F3}d}), Li ions flow in the system and lattice vibrations still carry a substantial amount of heat (see analysis below for details). More interestingly, the scattering between the flowing Li ions and the longitudinal acoustic modes is found to be much stronger than that occurring with the transverse acoustic modes. Transverse acoustic modes are related to the shear stress among ions and the longitudinal modes are the consequence of normal stress between two ions. When Li-ions flow, the shear stress between a specific Li-ion and other ions becomes weaker and thus the scattering of Li ions with transverse modes is weak compared to that on the longitudinal modes. By further increasing the system temperature to 1100 K, both Li and S ions enter the liquid state (\textbf{Figure ~\ref{fig:F2}e} and \textbf{\ref{fig:F2}f}). Although the scattering among different heat carriers, which are vibrations and ions’ flow, are quite strong, there are still propagating modes existing in liquid Li$_2$S (\textbf{Figure ~\ref{fig:F3}e} and \textbf{\ref{fig:F3}f}) and heat conduction largely contributes to the total thermal energy exchange. 

\begin{figure}
\hspace*{-4mm} 
\setlength{\abovecaptionskip}{0.1in}
\setlength{\belowcaptionskip}{-0.1in}
\includegraphics [width=3.5in]{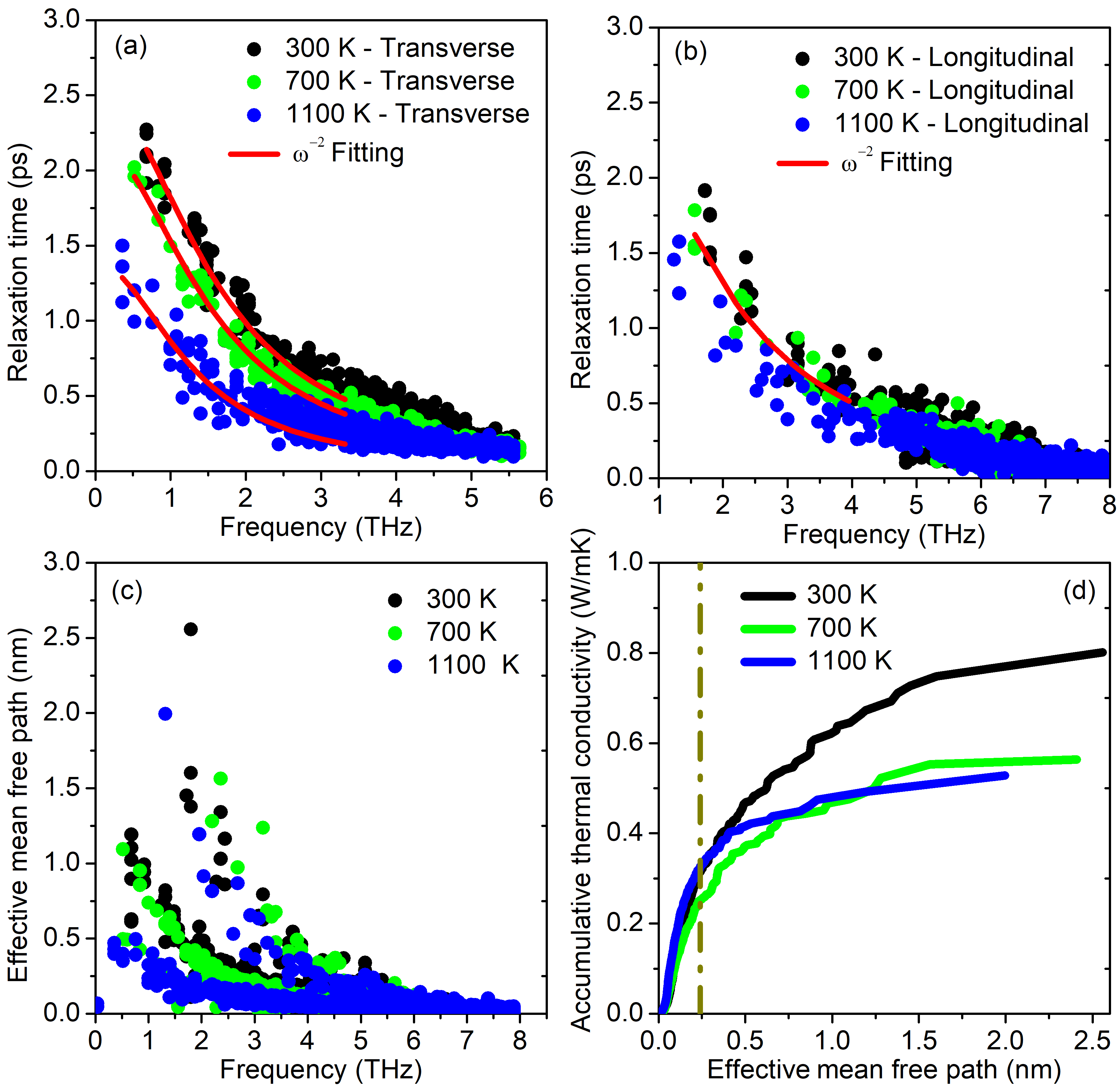}
\caption{Relaxation time of (a) the transverse branch and (b) the longitudinal branch at several temperatures. (c) Corresponding effective mean free path. (d) Accumulative thermal conductivity with respect to the effective mean free path for Li$_{2}$S at 300 K, 700 K and 1100 K.}
\label{fig:F4}
\end{figure}

We further calculate the mode relaxation time $\tau $ of vibrations by fitting the dynamical structure factor with the Lorentz function \cite{Zhou20152, Zhou2020, Thomas2010}
\begin{equation}
S=\frac{{{I}_{p}}}{{{\left[ (\omega -{{\omega }_{p}}\ )/\Delta  \right]}^{2}}+1}
\label{eqn:A6}  
\end{equation}
where ${{I}_{p}}$ and ${{\omega }_{p}}$ are the magnitude and the frequency at the peak center, respectively. The linewidth $\Delta $ yields half of the scattering rate $\Gamma $. The relaxation time can be then calculated as $\tau =1/2\Delta $ and follows a clear ${{\omega }^{-2}}$ scaling below 2.5 THz for the transverse branch (red line in \textbf{Figure ~\ref{fig:F4}a}) and below 4 THz for the longitudinal branch (red line in \textbf{Figure ~\ref{fig:F4}b}). The clear ${{\omega }^{-2}}$ scaling indicates that there exists propagating modes in the systems and their contribution to thermal conductivity is substantial (see analysis below for details). As discussed above, longitudinal vibrations undergo a larger number of scattering events and therefore the relaxation times of the transverse modes are generally larger than the ones of the longitudinal modes in the low-frequency region. With the increase of the system temperature, from 300 K to 700 K (1100 K), the relaxation time becomes smaller due to the increase of Umklapp scattering and to the interaction between vibrations and the Li-ions flow \cite{Zhou2018} since both Li and S ions flow at 1100 K (\textbf{Figure ~\ref{fig:F2}}). 

\vspace*{-2mm} 

We then move to the estimation of the effective mean free path ${{\Lambda }_{eff}}$ of the vibrational modes (\textbf{Figure ~\ref{fig:F4}c}). To calculate this quantity, we first estimate the group velocity ${{v}_{g}}=d\omega /dk$ using the fitted dispersions (black dashed lines in \textbf{Figure ~\ref{fig:F2}}) and write ${{\Lambda }_{eff}}={{v}_{g}}\cdot \tau $. As shown in \textbf{Figure ~\ref{fig:F4}c}, the effective mean free paths of a large amount of vibrations in the low-frequency region is larger than 0.24 nm corresponding to the distance between nearest neighboring ions and the effective mean free path of a fraction of the low-frequency vibrations can be as large as 2.5 nm at 300 K (2.2 nm at 700 K and 2.0 nm at 1100 K). As suggested by Figures 3a and 3b, those latter mean free paths confirm that a large population of the low-frequency vibrations consists of propagating modes that can be assimilated to propagons. The vibrations with effective mean free paths smaller than 0.24 nm cannot be treated as plane waves anymore since their wavelength cannot be defined on the atomic lattice. They are consequently categorized as diffusons in agreement with the theory of Allen and Feldman \cite{Allen1993}. Diffusons carry energy through harmonic coupling between non-propagating vibrations \cite{Allen1993, Moon2018} or equivalently through the overlap of the vibrations\rq \ \cite{Zhou2018} (as shown in \textbf{Figure ~\ref{fig:F2}}). The effective mean free path also decreases with the increase of the system temperature due to the enhancement of vibrational scattering as well as of the vibrations-ions\rq \  flow scattering. 

\begin{figure*}
\setlength{\abovecaptionskip}{0.1in}
\setlength{\belowcaptionskip}{-0.1in}
\includegraphics [width=5.5in]{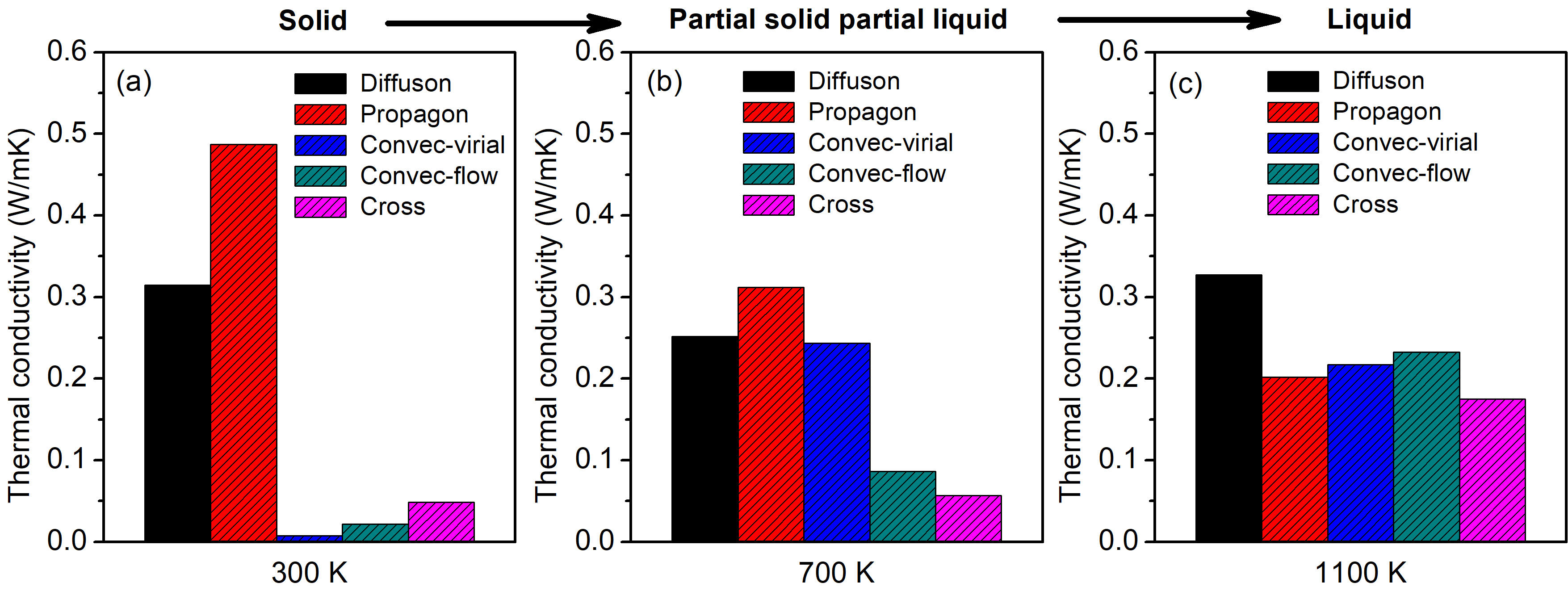}
\caption{Thermal conductivity contributed from each type of heat carrier in Li$_{2}$S in (a) the amorphous solid state (300 K), (b) the partial solid and partial liquid state (700 K) and (c) the liquid state (1100 K).}
\label{fig:F5}
\end{figure*}

To facilitate the analysis of thermal conductivity contributions, the thermal conductivity accumulation function ${{\kappa }_{vibration}}({{\Lambda }_{eff}})$ of vibrational modes has been computed from the effective mean free path ${{\Lambda }_{eff}}$ using
\begin{equation}
{{\kappa }_{vibration}}({{\Lambda }_{0}})=\sum\limits_{{{\Lambda }_{eff}}<{{\Lambda }_{0}}}{{{c}_{v}}{{v}_{g}}}{{\Lambda }_{eff}}=\sum\limits_{{{\Lambda }_{eff}}<{{\Lambda }_{0}}}{{{c}_{v}}v_{g}^{2}}\tau 
\label{eqn:A7}  
\end{equation}
where ${{c}_{v}}$ is the volumetric heat capacity which can be estimated by ${{c}_{v}}={{k}_{b}}/V$. Our results (Figure 4d) show that the contribution to thermal conductivity from vibrational heat is 0.80 W/mK at 300 K, 0.56 W/mK at 700 K, and 0.52 W/mK at 1100 K. It is not surprising that the thermal conductivity contributed from vibrations at 300 K is almost equal to the virial thermal conductivity, i.e 0.81 W/mK because Li$_2$S at 300 K is in its amorphous solid-state (\textbf{Figure ~\ref{fig:F1}a}, \textbf{Figure ~\ref{fig:F2}a} and \textbf{\ref{fig:F2}b}). The main thermal transfer inside the system occurs via conduction, which is resulting from the interatomic forces, i.e. the virial terms in Eq.\ (\ref{eqn:A1}). Contrastingly, the system at 700 K (1100 K) has a virial thermal conductivity of 0.79 W/mK (0.83 W/mK) much larger than the one contributed by vibrations. As discussed above (\textbf{Figure ~\ref{fig:F2}c} to \textbf{\ref{fig:F2}f}), Li (Li and S) ions do not vibrate around their equilibrium positions at 700 K (1100 K) and therefore the interatomic forces of the virial term not only contribute to conduction but also to convection. Particularly, our results show that diffusons (the dash-dotted line in \textbf{Figure ~\ref{fig:F4}d}) significantly contribute to the total thermal conductivity. 

Based on the above analysis, we now quantitively characterize the contribution to heat exchange from different thermal excitations, i.e. propagons ${{\kappa }_{propagon}}$, diffusons ${{\kappa }_{diffuson}}$, convection term due to interatomic forces ${{\kappa }_{convec\_virial}}$, convection term due to the ions\rq \  flow ${{\kappa }_{convec\_flow}}$ and, at last, the interaction between conduction and heat transfer ${{\kappa }_{cross}}$. Note that we estimate the contribution of the virial interatomic forces to convection${{\kappa }_{convec\_virial}}$ via ${{\kappa }_{convec\_virial}}={{\kappa }_{virial}}-{{\kappa }_{vibrations}}$. 

\begin{figure}[H]
\hspace*{-3mm} 
\setlength{\abovecaptionskip}{0.1in}
\setlength{\belowcaptionskip}{-0.1in}
\includegraphics [width=3.5in]{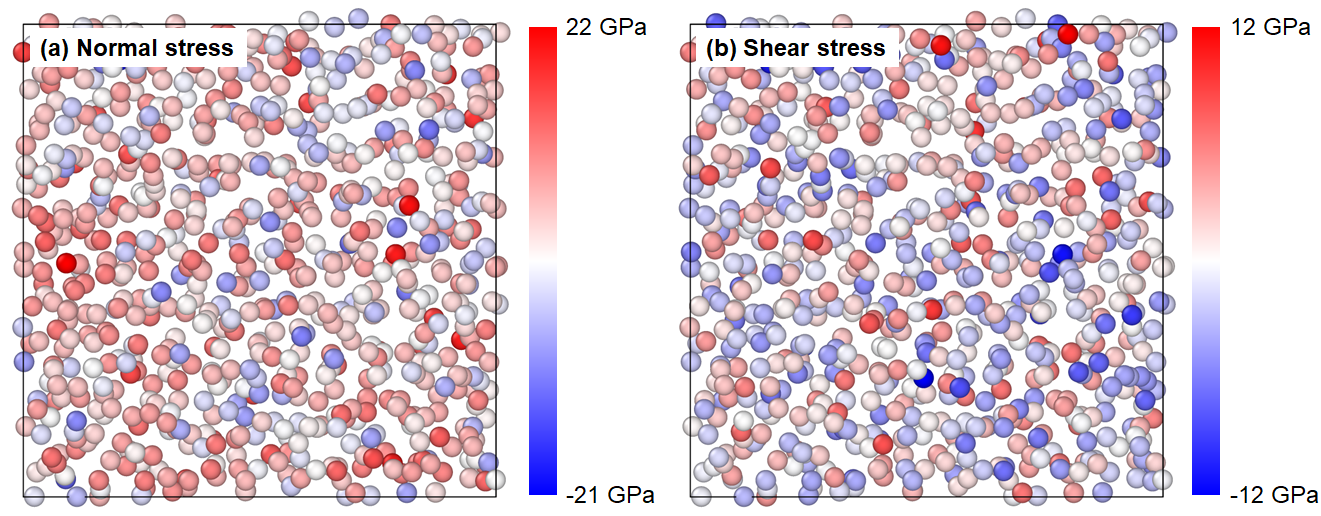}
\caption{Thermal conductivity contributed from each type of heat carrier in Li$_{2}$S in (a) the amorphous solid state (300 K), (b) the partial solid and partial liquid state (700 K) and (c) the liquid state (1100 K).}
\label{fig:F6}
\end{figure}

At 300 K (\textbf{Figure ~\ref{fig:F5}a} and Table 1), diffusons and propagons appears as the two main heat carriers in amorphous Li$_{2}$S, contributing to one-third and one half of the total thermal conductivity, respectively. The virial thermal conductivity is almost equal to the summed thermal conductivity of diffusons and propagnons. For the Li$_{2}$S system at 700 K (\textbf{Figure ~\ref{fig:F5}b} and \textbf{Tabel ~\ref{tab:table1}}), Li ions begin to flow and therefore the convection movement among the ions also carries one-third of the thermal energy. This convection thermal conductivity includes i) the thermal energy carried by the ions\rq \  flow and ii) the thermal energy transferred via virial interatomic interactions. The contribution from vibrations including propagons (0.31 W/mK) and diffusons (0.25 W/mK) becomes smaller because of the enhancement of the scattering among vibrations as well as of the one of the vibrations-ions’ flow. When both Li and S ions flow at 1100 K (\textbf{Figure ~\ref{fig:F5}c} and \textbf{Tabel ~\ref{tab:table1}}), the convection motion of ions contributes to half of the total thermal energy exchange and the thermal conductivity resulting from the vibrations remains large with a value of 0.53 W/mK. We also find that the cross term contributed from the interaction between conduction and convection reaches 0.24 W/mK and cannot be ignored in liquid Li$_{2}$S.
\begin{table*}
\caption{\label{tab:table1}Contributions to thermal transfer of each heat carrier in Li2S at 300 K, 700 K and 1100 K.}
\begin{ruledtabular}
\begin{tabular}{cccccc}
Temperature (K)&Diffusons (\%)&Propagons (\%)&Convection-interatomic (\%)&Convection-flow (\%)&Cross (\%)\\ \hline
300&0.362&	0.560&0.008&0.022&0.048 \\
700&0.275&	0.340&0.243&0.086&0.056\\
1100&0.233	&0.143&0.217&0.232&0.175\\
\end{tabular}
\end{ruledtabular}
\end{table*}
We now explain why propagating modes appear in liquid Li$_{2}$S and why convection contributes via the virial heat flux in PSPL and liquid Li$_{2}$S. In a simple liquid, longitudinal acoustic modes are always observed in experiments [18–20] because the density fluctuations can occur due to stronger repulsive forces between atoms when they move closer to each other. The transverse modes are usually hardly detectable in liquids because the shear force in the system is very weak. However, the situation changes when the vibrational effective mean free path in a liquid approaches the nearest neighbor distance between atoms. In this case, a solid-like cage effect at the nanometer scale acts as a restoring force for transverse vibrational modes \cite{Hosokawa2009}. To further confirm this proposition, we calculate the normal and shear stress in liquid Li$_{2}$S at 1100 K. Our results (\textbf{Figure ~\ref{fig:F6}}) show that the shear stress in the system reaches the same magnitude than the one of the normal stress, which indicates that the transverse modes in liquid Li$_{2}$S should also contribute largely to thermal transfer.

\vspace*{-1mm} 

We now turn to the expression of the virial heat flux which corresponds to the second and third terms in Eq.\ (\ref{eqn:A1}). In the solid, atoms or ions vibrate around their equilibrium positions and it can be assumed, at a reasonably low temperature, that ${{r}_{ij}}\approx r_{ij}^{0}$ where $r_{ij}^{0}$ is the distance between the equilibrium positions of atoms $i$ and $j$. In this situation, the thermal energy can be transferred only through conduction. When Li or S ions flow in the systems, ${{r}_{ij}}$ can significantly vary as shown in \textbf{Figure ~\ref{fig:F2}c}, \textbf{\ref{fig:F2}e} and \textbf{\ref{fig:F2}f}, which indicates that the virial heat flux can be also caused by the convection motion of atoms $i$ and $j$. 

In conclusion, by performing reaction force filed atomistic simulations and dynamic structure factor analysis, we show the map of the quantitative contributions to heat exchange of each thermal excitation in Li$_{2}$S from amorphous solid-state to partial-solid partial-liquid state and then to liquid-state. Our results reveal that the vibrations scattering time is in a wide-ranging from a few femtoseconds to several picoseconds for amorphous solid Li$_{2}$S at 300 K. As a result, both propagons and diffusons are the dominant heat carriers. While for the partial solid partial liquid Li$_{2}$S at 700 K, the vibrational scattering time is decreased comparing to that at 300 K, and the vibrational thermal conductivity decreases from 0.81 W/mK to 0.56 W/mK due to the enhancement of vibrational scattering and the scattering between vibrations and the Li ions flow. We also find that the thermal conductivity resulting from the convection heat transfer consists of i) the heat carried by the ions’ flow and ii) the heat induced by atomic interactions via the cross-correlation between the virial and the convection terms. For liquid Li$_{2}$S at 1100 K, the convection conductivity due to the cross-correlation is increased from a negligible value of 0.02 W/mK in solid Li$_{2}$S at 300 K to 0.63 W/mK at 1100K. The vibrational scattering time can be still as large as $\sim $ 1.5 ps. Thus, conduction resulting from both propagons and diffusons decreases largely compared to its weight in solid Li$_{2}$S, but remains quite substantial with a thermal conductivity value of 0.52 W/mK. Our results support a qualitatively different picture from the one of the state-of-the-art as we highlight that the virial heat flux also contributes to convection heat transfer and that propagating and non-propagating vibrational modes exist in liquid. Those latter significantly contribute to the thermal exchange by one third for liquid Li$_{2}$S at 1100 K. Our investigations provide a clear physical insight in the contributions of each thermal excitation at play in amorphous materials from solid to liquid.  

Y.Z. thanks startup fund from The Hong Kong University of Science and Technology (HKUST). This work used the Extreme Science and Engineering Discovery Environment (XSEDE), which is supported by the National Science Foundation grant number DMR180057.


\begin{thebibliography}{99}
\bibitem{Lan2010}
Y. Lan, A. J. Minnich, G. Chen, and Z. Ren, \href{https://doi.org/10.1002/adfm.200901512}{Adv. Funct. Mater. {\bf20}, 357 (2010).}
\bibitem{Bux2009}
S. K. Bux, R. G. Blair, P. K. Gogna, H. Lee, G. Chen, M. S. Dresselhaus, R. B. Kaner, and J. P. Fleurial, \href{https://doi.org/10.1002/adfm.200900250}{Adv. Funct. Mater. {\bf19}, 2445 (2009).}
\bibitem{Huang2019}
X. Huang, C. Zhu, Y. Lin, and G. Fang, \href{https://doi.org/10.1016/j.applthermaleng.2018.11.007}{Appl. Therm. Eng. {\bf147}, 841 (2019).}
\bibitem{Cahill2003}
D. G. Cahill, W. K. Ford, K. E. Goodson, G. D. Mahan, A. Majumdar, J. Humphrey, R. Merlin, S. R. Phillpot, W. K. Ford, and H. J. Maris, \href{https://doi.org/10.1063/1.1524305}{J. Appl. Phys. {\bf93}, 793 (2003).}
\bibitem{Kittel1976}
C. Kittel, $Introduction\ to\ Solid\ State\ Physics$ (Wiley, New York, 1976).
\bibitem{Allen1993}
P. B. Allen and J. L. Feldman, \href{https://doi.org/10.1103/PhysRevB.48.12581}{Phys. Rev. B {\bf48}, 12581 (1993).}
\bibitem{Moon2018}
J. Moon, B. Latour, and A. J. Minnich, \href{https://doi.org/10.1103/PhysRevB.97.024201}{Phys. Rev. B {\bf97}, 024201 (2018).}
\bibitem{Larkin2014}
J. M. Larkin and A. J. H. McGaughey, \href{https://doi.org/10.1103/PhysRevB.89.144303}{Phys. Rev. B {\bf89}, 144303 (2014).}
\bibitem{Zhou2015}
Y. Zhou and M. Hu, \href{https://doi.org/10.1103/PhysRevB.92.195205}{Phys. Rev. B {\bf92}, 195205 (2015).}
\bibitem{Saaskilahti2016}
K. S\"a\"askilahti, J. Oksanen, J. Tulkki, S. Volz, J. and A. J. H. Mcgaughey, \href{https://doi.org/10.1063/1.4968617}{AIP Adv. {\bf6}, 121904 (2016).}
\bibitem{Zhou2016}
Y. Zhou and M. Hu, \href{https://doi.org/10.1021/acs.nanolett.6b02450}{Nano Lett. {\bf16}, 6178 (2016).}
\bibitem{Cahill1992}
D. G.Cahill, \href{https://doi.org/10.1103/PhysRevB.46.6131}{Phys. Rev. B {\bf46}, 6131 (1992).}
\bibitem{Lv2016}
W. Lv and A. Henry, \href{https://doi.org/10.1088/1367-2630/18/1/013028}{New J. Phys. {\bf18}, 013028 (2016).}
\bibitem{Zhou2018}
Y. Zhou, S. Xiong, X. Zhang, S. Volz, and M. Hu, \href{https://doi.org/10.1038/s41467-018-07027-x}{Nat. Commun. {\bf9}, 4712 (2018).}
\bibitem{Wu2018}
B. Wu, Y. Zhou, and M. Hu, \href{https://doi.org/10.1021/acs.jpclett.8b02542}{J. Phys. Chem. Lett. {\bf9}, 5704 (2018).}
\bibitem{Li2018}
B. Li, H. Wang, Y. Kawakita, Q. Zhang, M. Feygenson, H. L. Yu, D. Wu, K. Ohara, T. Kikuchi, K. Shibata, T. Yamada, X. K. Ning, Y. Chen, J. Q. He, D. Vaknin, R. Q. Wu, K. Nakajima, and M. G. Kanatzidis, \href{https://doi.org/10.1016/j.nanoen.2015.10.002}{Nat. Mater. {\bf17}, 226 (2018).}
\bibitem{Qiu2014}
W. Qiu, L. Xi, P. Wei, X. Ke, J. Yang, and W. Zhang, \href{https://doi.org/10.1016/j.nanoen.2017.06.021}{Proc. Natl. Acad. Sci. {\bf111}, 15031 (2014).}
\bibitem{Hosokawa2009}
S. Hosokawa, M. Inui, Y. Kajihara, K. Matsuda, T. Ichitsubo, W. C. Pilgrim, H. Sinn, L. E. González, D. J. González, S. Tsutsui, and A. Q. R. Baron, \href{https://doi.org/10.1103/PhysRevLett.102.105502}{Phys. Rev. Lett. {\bf102}, 105502 (2009).}
\bibitem{Elton2016}
D. C. Elton and M. Fernández-Serra, \href{https://doi.org/10.1038/ncomms10193}{Nat. Commun. {\bf7}, 10193 (2016).}
\bibitem{Woods1970}
A. D. B. Woods and R. A. Cowley,  \href{https://doi.org/10.1103/PhysRevLett.24.646}{Phys. Rev. Lett. {\bf24}, 646 (1970).}
\bibitem{SM2020}
See Supplemental Material for details of the size effect and the convergence of the Green-Kubo thermal conductivity.
\bibitem{Islam2015}
M. M. Islam, A. Ostadhossein, O. Borodin, A. T. Yeates, W. W. Tipton, R. G. Hennig, N. Kumar, and A. C. T. Van Duin, \href{https://doi.org/10.1039/C4CP04532G}{Phys. Chem. Chem. Phys. {\bf17}, 3383 (2015).}
\bibitem{Kubo1957}
R. Kubo, \href{https://doi.org/10.1143/JPSJ.12.570}{J. Phys. Soc. Japan {\bf12}, 570 (1957).}
\bibitem{Hardy1963}
R. J. Hardy, \href{https://doi.org/10.1103/PhysRev.132.168}{Phys. Rev. {\bf132}, 168 (1963).}
\bibitem{Shintani2008}
H. Shintani and H. Tanaka, \href{https://doi.org/10.1038/nmat2293}{Nat. Mater. {\bf7}, 870 (2008).}
\bibitem{Zhou20152}
Y. Zhou, X. Zhang, and M. Hu,\href{https://doi.org/10.1103/PhysRevB.92.195204}{Phys. Rev. B {\bf92}, 195204 (2015).}
\bibitem{Zhou2020}
Y. Zhou, J. Tranchida, Y. Ge, J. Y. Murthy and T. S. Fisher, \href{https://arxiv.org/abs/1901.00966}{ArXiv:1901.00966 (2019)}
\bibitem{Thomas2010}
J. A. Thomas, J. E. Turney, R. M. Iutzi, C. H. Amon, and A. J. H. McGaughey,\href{https://doi.org/10.1103/PhysRevB.81.081411}{Phys. Rev. B {\bf81}, 081411(R) (2010).}
\end{thebibliography}
\end{document}